\documentclass{iopart}
\usepackage{graphicx}
\usepackage{epsfig}
\def\lsim{\raise0.3ex\hbox{$<$\kern-0.75em\raise-1.1ex\hbox{$\sim$}}}
\def\gsim{\raise0.3ex\hbox{$>$\kern-0.75em\raise-1.1ex\hbox{$\sim$}}}
\begin{document}

\title[News from Lattice QCD]{News from Lattice QCD on Heavy Quark
Potentials and Spectral Functions of Heavy Quark States
%\vskip -120pt
%\mbox{} \hfill BI-TP 04/xx\\
%\mbox{} \hfill March 2004\\
%\vskip 85pt
} 

\author{Frithjof Karsch\footnote[3]{karsch@physik.uni-bielefeld.de}
}
\address{Fakult\"at f\"ur Physik, Universit\"at Bielefeld, D-33615
Bielefeld, Germany} 

\begin{abstract}
We discuss recent lattice results on in-medium properties of hadrons
and focus on thermal properties of heavy quark bound states. We
will clarify the relation between heavy quark free energies and 
potentials used to analyze the melting of heavy quark bound states.
Furthermore, we present calculations of meson spectral functions 
which indicate that the charmonium ground states, $J/\psi$ and $\eta_c$,
persist in the quark gluon plasma as well defined resonances with no 
significant change of their 
zero temperature masses at least up to $T\simeq 1.5 T_c$. We also briefly
comment on the current status of lattice calculations at non-vanishing
baryon number density.
\end{abstract}

%Uncomment for PACS numbers title message
%\pacs{00.00, 20.00, 42.10}

% Uncomment for Submitted to journal title message
%\submitto{\JPA}

% Comment out if separate title page not required
%\maketitle

\section{Introduction} 

Two quite different aspects of the studies of QCD thermodynamics on the lattice
gained most attention in the heavy ion community during recent years -- 
%thermodynamics of strongly interacting
%matter have been in the focus of studies of lattice regularized QCD 
%during the recent years -- 
studies of the QCD phase diagram at non-zero baryon chemical potential and 
the analysis of the influence of a thermal heat bath of (quarks and)
gluons on basic properties of hadrons, e.g. their masses and widths. 
In this review we will focus on the
latter aspect of lattice studies but will also briefly discuss studies of
the QCD phase diagram in this Introduction. 

\subsection{Finite density QCD}

There has been 
considerable progress in extending studies of the QCD phase diagram to 
small but non-zero values of the baryon chemical potential $\mu_B$
\cite{Fodor1,Allton1,Philipsen1}. This led to a first analysis of
the equation of state at non-vanishing baryon number density 
\cite{Fodor2,Allton2,Gavai1}
and estimates for the location of the chiral critical point (second order phase
transition) \cite{Fodor1} at which the transition to the high temperature 
phase of QCD turns from a rapid crossover to a first order phase
transition. Recent calculations suggest
a critical value $\mu_B^{crit} \sim 400$~MeV \cite{Allton3,Fodor3}.
This is much larger than the estimate for the baryon chemical 
potential which characterizes the chemical freeze-out of hadron
resonances at RHIC, $\mu_B^{freeze} \simeq 29$~MeV \cite{PBM} and
suggests that the transition at $\mu_B = 0$ as
well as in the almost baryon free region produced at mid-rapidity at
RHIC is characterized by a rapid crossover from a hadronic resonance 
gas to the quark gluon plasma. The analysis of the equation of state for
$\mu_B \; \ge\; 0$ shows,
on the one hand, that thermodynamics in the low temperature phase of QCD
is well approximated by thermal properties of a hadronic resonance
gas \cite{Redlich}. On the other hand, it also shows that the strong
increase of baryon number fluctuations, which for 
$\mu_B > 0$ and $T \;<\; T_c$ is consistent with the rise of baryon number
fluctuations in a resonance gas, is suppressed once the system enters
the plasma phase \cite{Allton2}. The resulting pronounced peak in the
baryon number fluctuations indicates the proximity of the chiral critical point 
in the QCD phase diagram. Although it has been suggested since a long
time that these kind of fluctuations should result also in visible 
event by event fluctuations in heavy ion collisions, it remains to be seen 
to what extent the experimentally observed increase in the $\displaystyle{
K^{+}/ \pi^{+}}$ ratio
\cite{horn} and its fluctuation \cite{Kpifluct} can be attributed to
the QCD phase transition.

\subsection{In-medium properties of hadrons}

Confinement and chiral symmetry breaking are main characteristics
of QCD which explain the basic properties of the hadron spectrum in
the vacuum. Qualitative features like the existence of a light
Goldstone particle, the pion, and the excitation spectrum of heavy
quark bound states can be understood through the mere
existence of a non-vanishing, spontaneously generated chiral condensate 
$\langle \bar{\psi}\psi \rangle$, and the linearly rising heavy quark
potential with a non-vanishing string tension $\sigma$, respectively. 
Quantitative properties of the spectrum depend, however, on the actual 
value and $T$-dependence of $\langle \bar{\psi}\psi\rangle$ and $\sigma$ 
as well as on other properties of QCD like the running of the coupling  
with temperature or the generation of a thermal, 
Debye screening mass. It recently has been pointed out that despite
of chiral symmetry restoration and deconfinement it is well conceivable
that a complicated quasi-particle structure may persist in the QCD plasma
phase at temperatures $T\; \gsim\; T_c$ \cite{shuryak}. In fact, the 
non-perturbative structure of the plasma phase for $T\; \lsim\; 2 T_c$ 
is already evident from the strong deviation of bulk thermodynamic 
quantities , e.g. the pressure, from their perturbative form \cite{Peikert}. 

\begin{figure}
\begin{center}
\hspace*{-2mm}
\epsfig{file=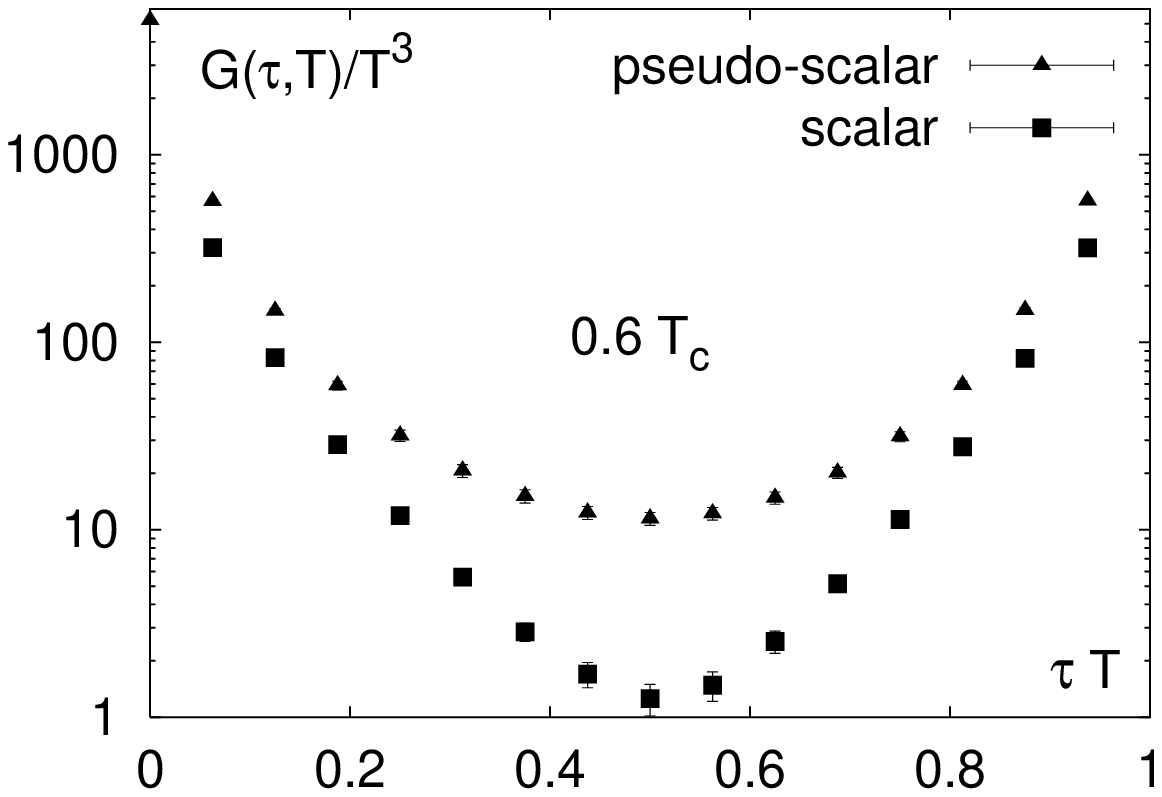,width=65mm}
\epsfig{file=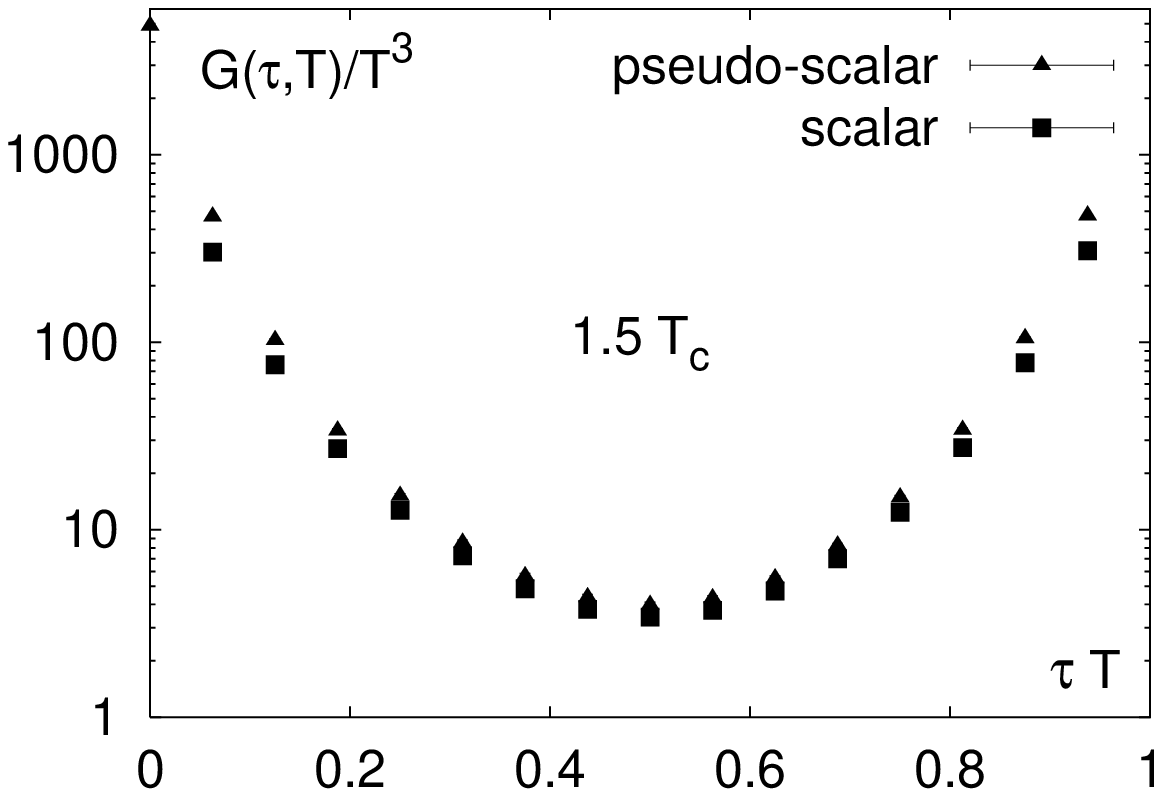,width=65mm}
\end{center}
\caption{\label{chiral} Scalar and pseudo-scalar correlation
functions at $T=0.6 T_c$ (left) and $T=1.5 T_c$ (right). Shown are results
from calculations with light quarks on a $64^3\times 16$ lattice in 
quenched QCD. 
}
\end{figure}

The influence of a thermal heat bath on hadron properties immediately
becomes evident when one compares Euclidean time ($\tau$) correlation 
functions of mesons
in different quantum number channels. At low temperature chiral symmetry 
breaking and the breaking of the axial $U(1)$ symmetry lead to a splitting  
of hadronic states, which would be degenerate otherwise and  
would then also have identical correlation functions. In
\Fref{chiral} we show pseudo-scalar ($\pi$) and scalar ($\delta$) meson
correlation functions in a gluonic heat bath (quenched QCD) at temperatures 
below (left) and above (right) the deconfinement transition temperature;
\begin{equation}
G_H(\tau, T) = \frac{1}{V} \sum_{\vec{x}}
\langle J_H(\tau, \vec{x}) J_H^\dagger (0, \vec{0}) \rangle \quad ,
\label{def2pt}
\end{equation}
with $J_H$ denoting a hadron current with the appropriate quantum
numbers of the hadronic channel $H$. The spectrum in the pseudo-scalar and
scalar channels differs at low temperature because of the explicit breaking
of the axial $U(1)$ symmetry which consequently leads to quite different 
correlation functions in \Fref{chiral}(left). 
The almost perfect degeneracy of both correlation functions above $T_c$, however,
suggests that this symmetry is effectively restored in the 
deconfined phase of QCD (\Fref{chiral}(right)).

The hadron correlation functions, $G_H(\tau,T)$, are directly related to 
spectral functions $\sigma_H(\omega, T)$ which contain all the information
on thermal modifications of the hadron spectrum in certain channels of
hadron quantum numbers, $H$,
\begin{equation}
G_H (\tau,T)
= \int_{0}^{\infty} {\rm d}\omega\;
\sigma_H (\omega, T) \;
{\cosh (\omega (\tau - 1/2T)) \over \sinh ( \omega /2T )} \quad .
\label{correlator}
\end{equation}
Lattice studies of in-medium properties of hadrons greatly advanced
in their predictive power through the exploitation of the Maximum
Entropy Method (MEM) \cite{Hatsuda1}. This
allows the reconstruction of $\sigma_H (\omega, T)$ at non-zero
temperature for light as well as heavy quark bound states. 
We will concentrate here on the heavy quark sector and, in particular,
will discuss in how far potential model calculations, which generally led 
to an early dissociation of heavy quark bound states \cite{Digal} can be
made consistent with recent studies of charmonium spectral
functions in the high temperature phase of QCD \cite{Asakawa1,Datta1,Umeda1},
which find charmonium bound states even at $T\simeq 1.5~T_c$. 
This will require to reconsider the relation between heavy quark
free energies calculated on the lattice and temperature dependent
potentials used in potential model calculations. We will start with
a discussion of this problem in the next section and present recent results
on charmonium spectral functions in section 3.

\section{Color averaged heavy quark free energies and color singlet
potentials}

\begin{figure}
\begin{center}
\hspace*{-2mm} 
\epsfig{file=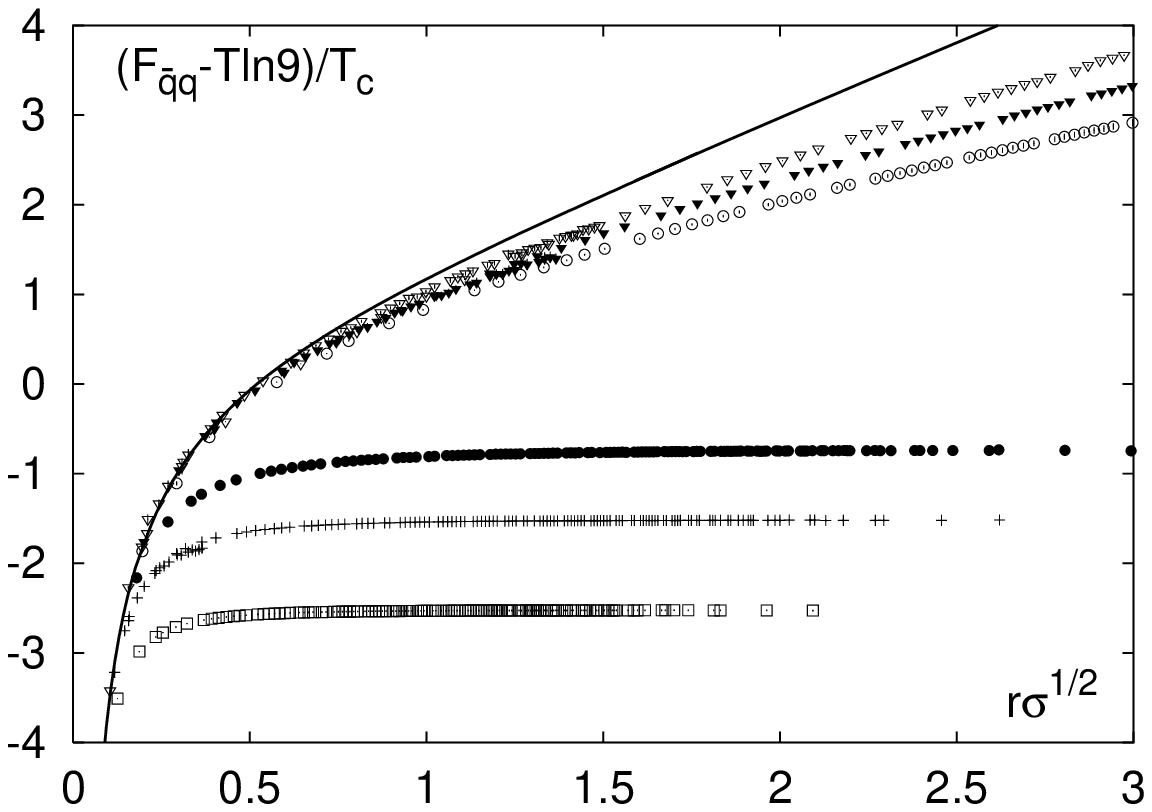,width=65mm}
\epsfig{file=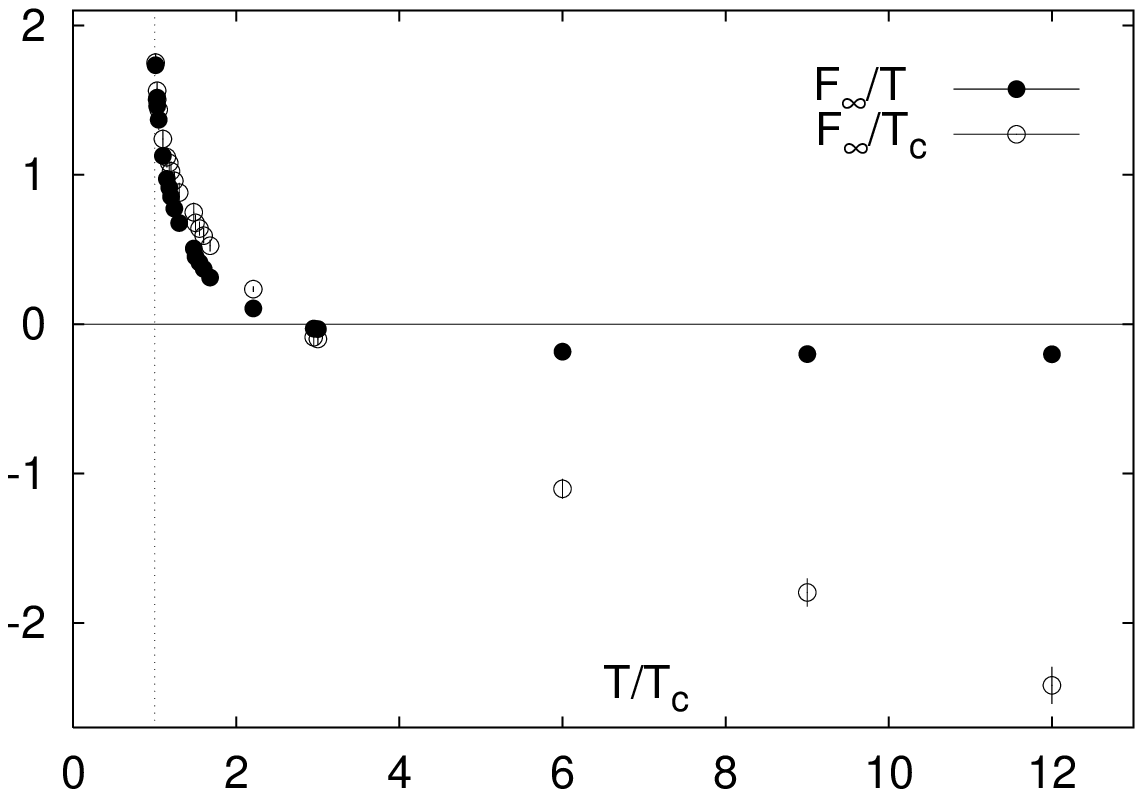,width=65mm}
\end{center}
\caption{\label{coloraveraged}The color averaged heavy quark free energy
at several values of the temperature (left) and the temperature dependence 
of its asymptotic value at large distances (right): The left hand figure
shows results for $T/T_c = 0.9,~0.94,~0.98,~1.05,~1.2,~1.5$ (from top to
bottom) obtained in quenched QCD.}
\end{figure}

The original idea of charmonium suppression as a signature for 
plasma formation in heavy ion collisions \cite{Matsui} relies 
on the occurrence of fundamental changes in the heavy quark potential 
(vanishing string tension, Debye screening) in the high temperature phase
of QCD. While there is no doubt that this does happen in QCD
it was clear right from the beginning of studies of 
$J/\psi$-suppression in potential models \cite{Mehr} that 
a quantitative determination of dissociation temperatures 
depends on the detailed structure of the heavy quark potential 
at high temperature. Information on the latter can be extracted
from an analysis of heavy quark free energies \cite{McLerran}
calculated on the lattice which describe the change in free 
energy of a thermal medium due to the presence of static quark
and anti-quark sources,   
\begin{equation}
{\rm e}^{-\; F_{\bar{q}q}(r,T)} = \frac{Z_{\bar{q}q} (T,V)}{Z (T,V)}
\equiv \frac{1}{V} \sum_{\vec{x},\vec{y}; |\vec{x}-\vec{y}|=r}
\frac{1}{9}
\langle {\rm Tr} L(\vec{x}){\rm Tr}L^{\dagger}(\vec{y})\rangle \quad . 
\label{freedef}
\end{equation}
Here the static sources are represented by Polyakov loops
$L(\vec{x})$ and $L^{\dagger}(\vec{y})$, and $Z_{\bar{q}q} (T,V),~Z (T,V)$
denote the QCD partition functions in presence or absence of these
sources, respectively. Quite often $F_{\bar{q}q}(r,T)$ itself has been 
interpreted
as the heavy quark potential at finite temperature and as such has been
used in potential model calculations. This, however, ignores the fact
that aside from modifying the energy of the thermal heat bath 
the external sources also change the entropy of the system.
The change of energy and entropy will depend on temperature
as well as the separation of the $\bar{q}q$-pair,
\begin{equation}
U_{\bar{q}q}(r,T) = T^2 \; \frac{\partial F_{\bar{q}q}(r,T)/T}{\partial T}
\quad , \quad    S_{\bar{q}q}(r,T) = - \frac{\partial
F_{\bar{q}q}(r,T)}{\partial T} \quad .
\label{US}
\end{equation}
The relative importance of energy and entropy contributions to the 
free energy at different $\bar{q}q$ separations becomes evident after
a proper renormalization of Polyakov loops \cite{Zantow,Pisarski}.
When renormalizing $L(\vec{x})$ by matching $\tilde{F}_{\bar{q}q}(r,T) =
F_{\bar{q}q}(r,T) -T\ln 9$ at short distances to the zero temperature 
heavy quark potential \cite{Zantow} one readily finds that at 
short distances the temperature dependence of $\tilde{F}$ is weak. This 
indicates that after subtraction of a simple statistical entropy
term ($T\; \ln 9$) the short distance part of the free energy reflects
the change in energy due to the presence of static quark anti-quark sources. 
This, however, is not sufficient to obtain the (potential) energy at all
distances. At large distances a strong temperature dependence remains
even after the subtraction of $T\ln 9$. In fact,
$\tilde{F}_{\bar{q}q}(r,T)$ and
$F_{\bar{q}q}(r,T)$ will approach $(-\; \infty)$ for $T\rightarrow
\infty$ (see \Fref{coloraveraged}a). As seen in
\Fref{coloraveraged}b, $F_\infty (T) \equiv F_{\bar{q}q}(\infty, T)$ decreases
almost linearly in $T$, {\it i.e.} $F_\infty (T) \sim -0.2 T$. This
shows that the free energies
contain a substantial entropy contribution at large distances while the
energy contribution almost vanishes, which according to \Eref{US} is 
reflected by the weak temperature dependence of $F_\infty (T)/T$. This
complicated $r$-dependent entropy contribution makes a direct use of
free energies in potential models questionable.

\begin{figure}
\begin{center}
\hspace*{-2mm}
\epsfig{file=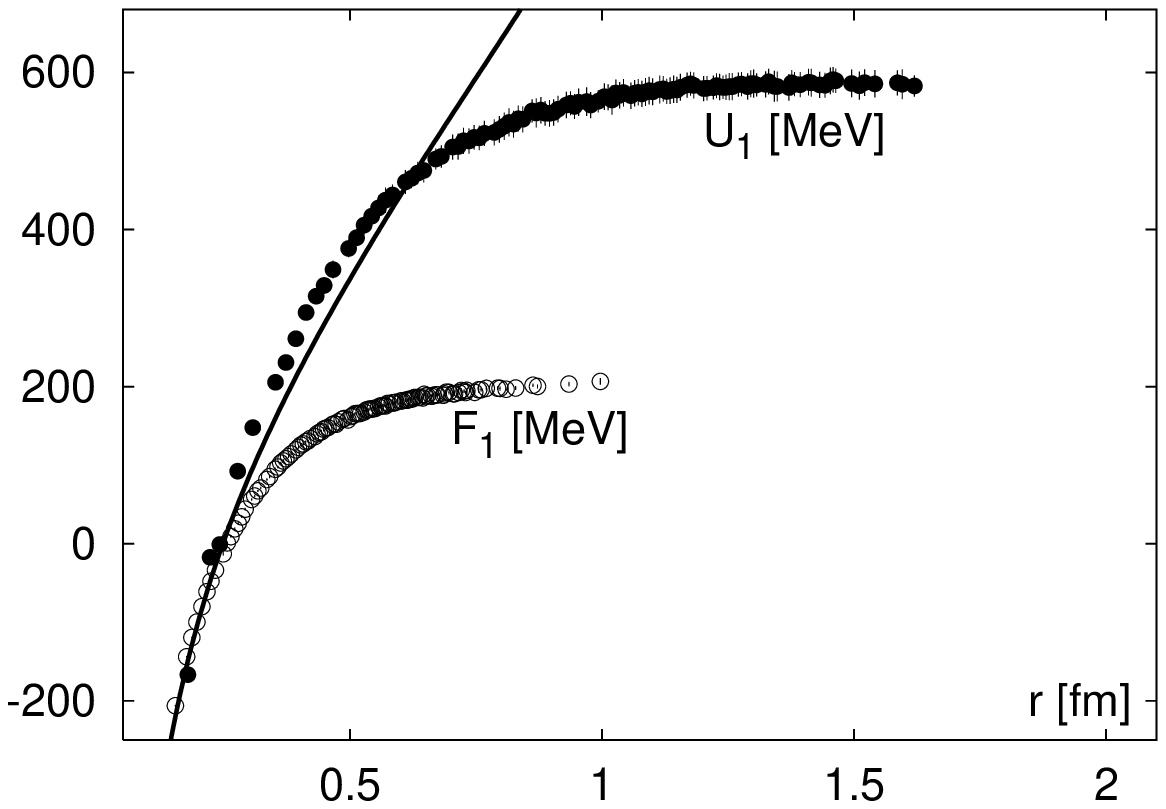,width=65mm}
\epsfig{file=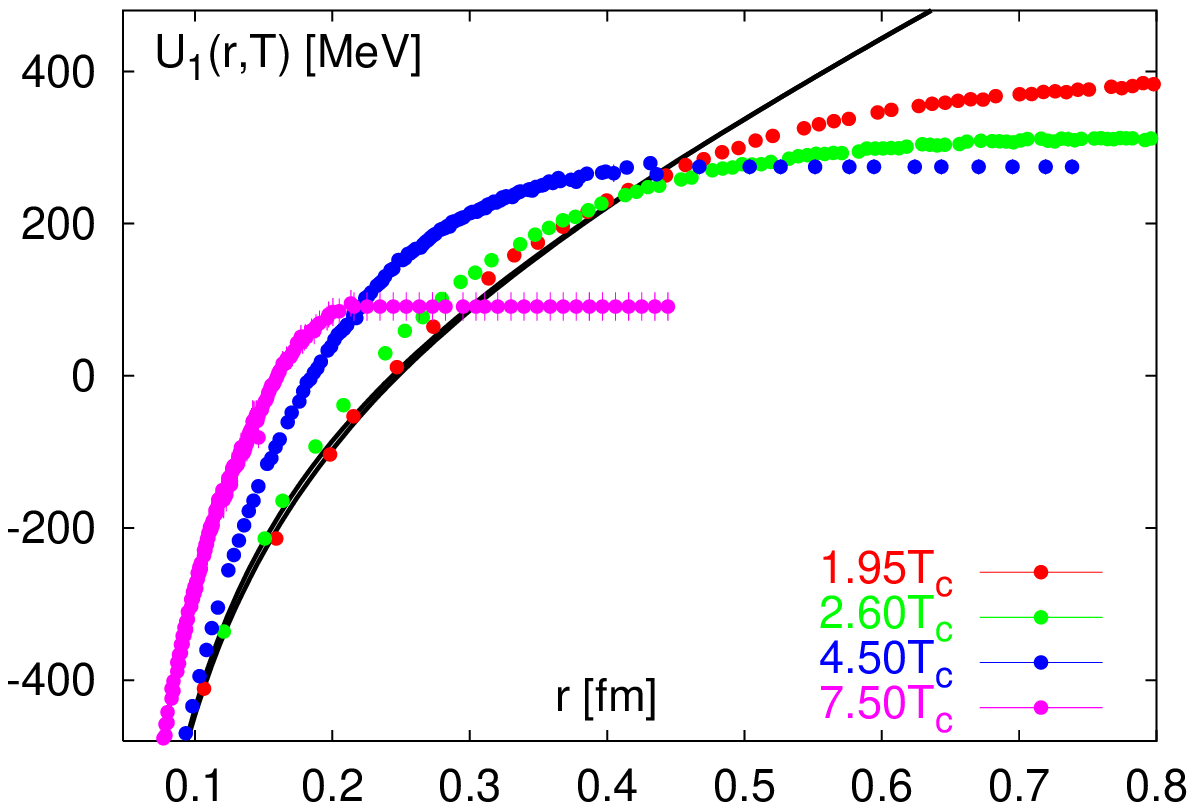,width=65mm}
\end{center}
\caption{\label{freeenergy}Singlet free energy and heavy quark energy 
in quenched QCD
calculated in Coulomb gauge at $T = 1.5\; T_c$ (left) and 
heavy quark energy at several other values of the temperature (right).}
\end{figure}

A further obstacle for a direct use of lattice calculations of heavy 
quark free energies in potential models arises because at non zero 
temperature $F_{\bar{q}q}$ does not directly give the singlet free energy
from which one could determine the singlet energy or potential used in 
phenomenological approaches.
The free energy defined through the Polyakov loop correlation function
in \Eref{freedef} gives a weighted average over singlet and octet contributions 
\cite{McLerran}. At non-zero temperature the definition of a singlet free 
energy requires gauge fixing. At present most studies 
of the singlet free energy have been performed in Coulomb gauge\footnote{To 
avoid an explicitly gauge dependent formulation one could introduce 
additional spatial 
Wilson lines to connect the quark anti-quark sources (cyclic Wilson loop). 
This trades ''gauge dependence'' for ''path dependence'' \cite{Nadkarni}.
However, also the singlet free energy calculated in Coulomb gauge can be
understood in terms of an appropriately defined gauge invariant operator
\cite{Philipsen}.}, where the singlet free energy is given by,
\begin{equation}
{\rm e}^{-\; F_{1}(r,T)} = 
\frac{1}{V} \sum_{\vec{x},\vec{y}; |\vec{x}-\vec{y}|=r}
\frac{1}{3}\langle {\rm Tr}L(\vec{x})L^{\dagger}(\vec{y})\rangle \quad . 
\label{freedef_1}
\end{equation} 
Some results for $T\; =\; 1.5 \; T_c$ are shown in \Fref{freeenergy}.
As expected from the discussion given above it is clear that at short
distances free energy and energy will coincide while at large distances
the asymptotic value of the heavy quark energy will be larger than the
corresponding free energy value as the entropy contribution has been
removed. This suggests that heavy quark bound states could survive  
in the QCD plasma up to temperatures significantly higher than those
estimated  previously in potential model
calculations, which have been based on input obtained from calculations
of the heavy quark free energy. To quantify these statements it is 
necessary to extent the analysis discussed here in the context of quenched
QCD also to the case of QCD with a realistic light quark spectrum.

\section{Spectral analysis of hadronic correlators at high temperature}

It should have become clear from the discussion in the previous section
that an analysis of thermal properties of heavy quark bound states in 
terms of potential models and probably even the determination of singlet
potentials from lattice calculations will require additional phenomenological 
consideration. On the other hand, as pointed out in the Introduction,
an ab-initio approach to thermal hadron properties through lattice 
calculations exists and is based on the calculation of thermal 
hadron correlation functions, $G_H(\tau,T)$. As indicated in \Eref{correlator}
$G_H(\tau,T)$ is  
related to the spectral functions $\sigma_H(\omega, T)$. However,
an inversion of this integral equation is generally not possible because
lattice calculations only yield information on $G_H(\tau,T)$ at a finite,
discrete set of Euclidean time steps, $\tau_k T = k/ N_\tau$ with $k=0, 1, ...
N_\tau -1$, with $N_\tau$ denoting the temporal extent of the lattice. 
It is, however, possible to determine the {\it most probable} spectral
function which describes the calculated data set 
$\{ G_H (\tau_k,T)\; | \; k=0, ..., N_\tau-1 \}$
and respects known constraints on $\sigma_H (\omega, T)$ 
(positivity, asymptotic behavior, ...).
This can be achieved using a Bayesian data analysis, e.g. the Maximum
Entropy Method (MEM) \cite{Bryan}. In the context of QCD calculations
this has been introduced in \cite{Hatsuda1}.

\begin{figure}
\begin{center}
\hspace*{-2mm}
\epsfig{file=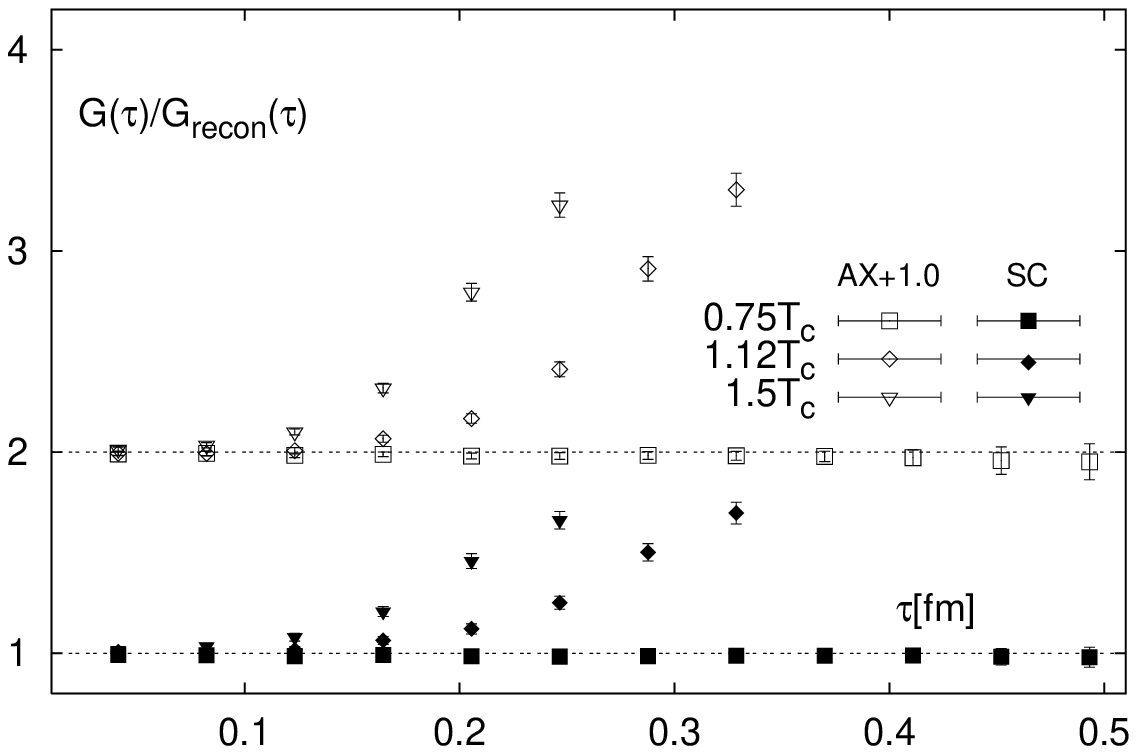,width=65mm}
\epsfig{file=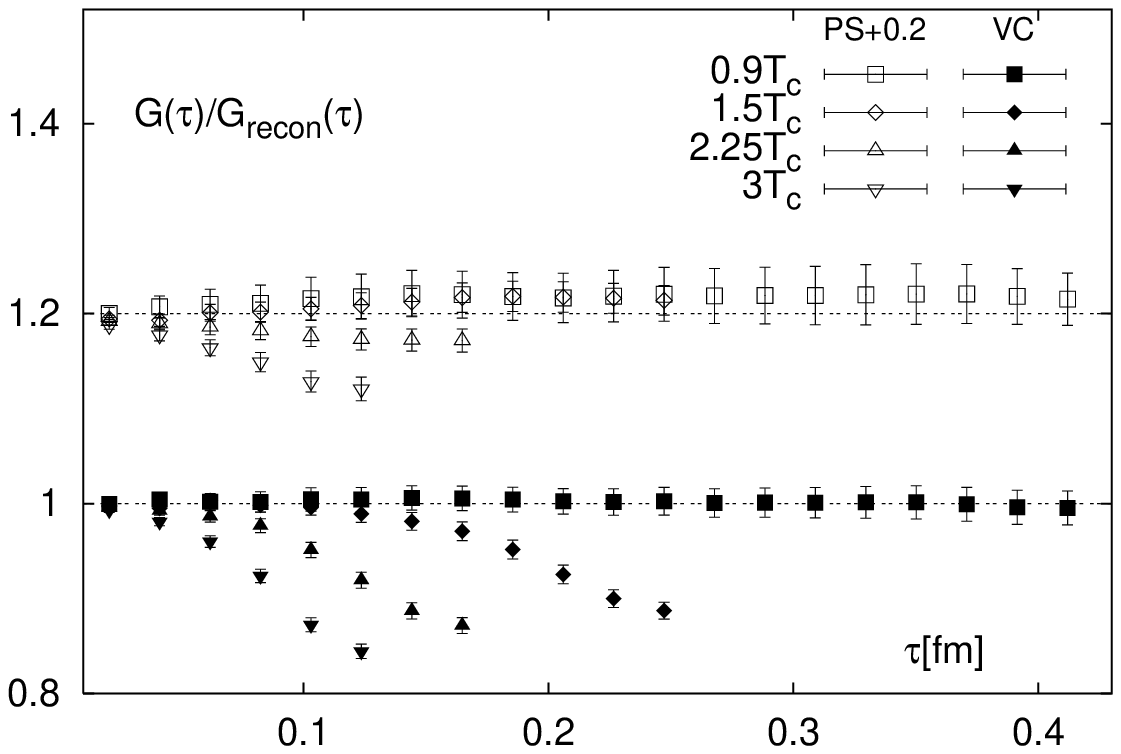,width=65mm}
\end{center}
\caption{\label{ratios}Thermal correlation functions above $T_c$ in axial 
vector (AX) scalar (SC), pseudo-scalar (PS) and vector (VC) channels normalized 
to reconstructed correlation functions which are based on spectral functions 
calculated at $T^{*} = 0.75 T_c$ (left) and $T^{*} = 0.9 T_c$ (right). Note
the different scales on the ordinates of both figures.}
\end{figure}

A first feeling for the influence of a thermal heat bath on the
structure of correlation functions and the size of medium modifications 
of a thermal spectral function can be obtained
by comparing directly the numerically calculated correlation functions
at temperature $T$ with a correlation function, $G_{\rm recon} (\tau_k,T)$,
constructed from the spectral function calculated at a smaller (zero) 
temperature, $\sigma_H (\omega , T^{*} )$ with $T^{*} < T$, {\it i.e.}
\begin{equation}
G_{\rm recon} (\tau,T)
= \int_{0}^{\infty} {\rm d}\omega\;
\sigma_H (\omega, T^{*}) \;
{\cosh (\omega (\tau - 1/2T)) \over \sinh ( \omega /2T )} \quad .
\label{correlator2}
\end{equation}
In this way the trivial temperature dependence of the integration kernel
is taken care of and any remaining discrepancies between the reconstructed
correlation function and the actually calculated correlation function at
temperature $T$ can be attributed to changes in the spectral function.
Some results for charmonium correlation functions are shown in \Fref{ratios} 
\cite{Datta1}. 

The ratio of correlation functions shown in \Fref{ratios} would equal unity, if 
the spectral functions would not depend on temperature. However, as can be seen 
the correlation functions for $P$-state charmonia (left) show strong thermal
modifications already at temperatures slightly above $T_c$. On the
other hand, the $S$-state correlators (right) show only little modifications up 
to $T\simeq 1.5\; T_c$. In fact,
no temperature dependence is visible in the pseudo-scalar channel ($\eta_c$)
up to $T= 1.5\; T_c$ while modifications in the vector channel ($J/\psi$) are
of the order of $10\%$ at this temperature. 

\begin{figure}
\begin{center}
\setlength{\unitlength}{1.0cm}
\begin{picture}(10.0,8.0)
\boldmath
\put(4.3,5.11){\epsfig{bbllx=0,bblly=175,bburx=564,bbury=514,
file=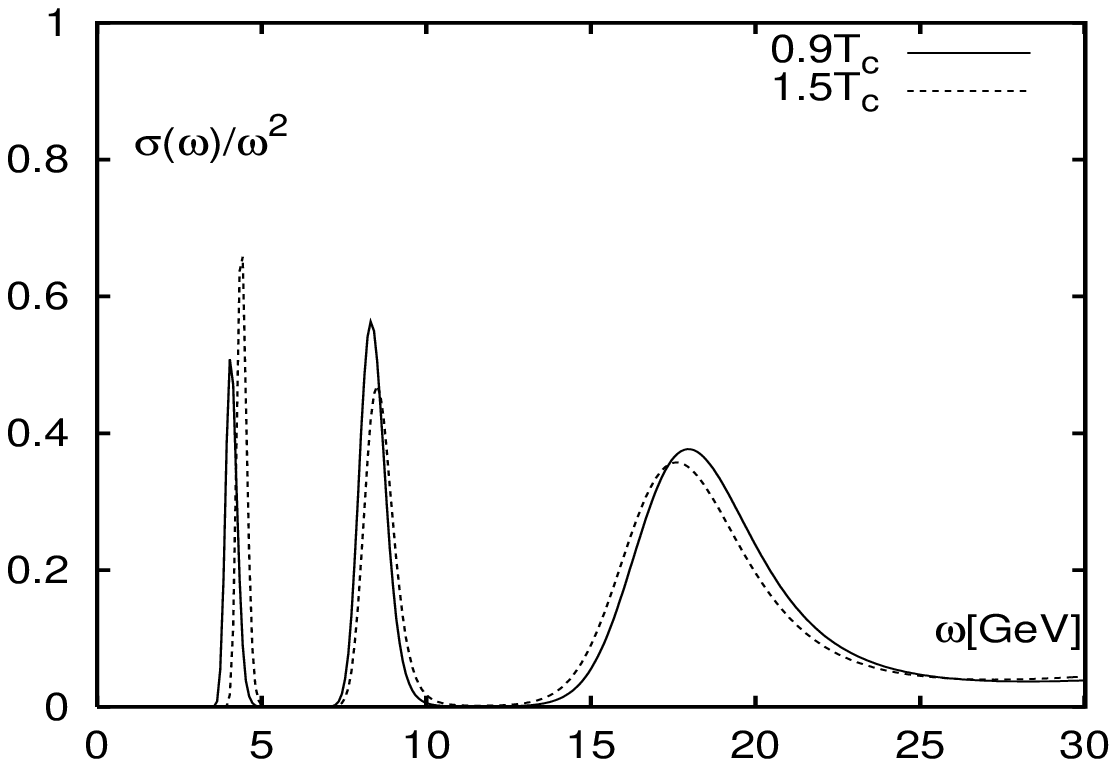,width=105mm}}
\put(4.2,2.77){\epsfig{file=whitebox2.eps,width=120mm,height=0.4cm}}
\put(4.3,1.41){\epsfig{bbllx=0,bblly=175,bburx=564,bbury=514,
file=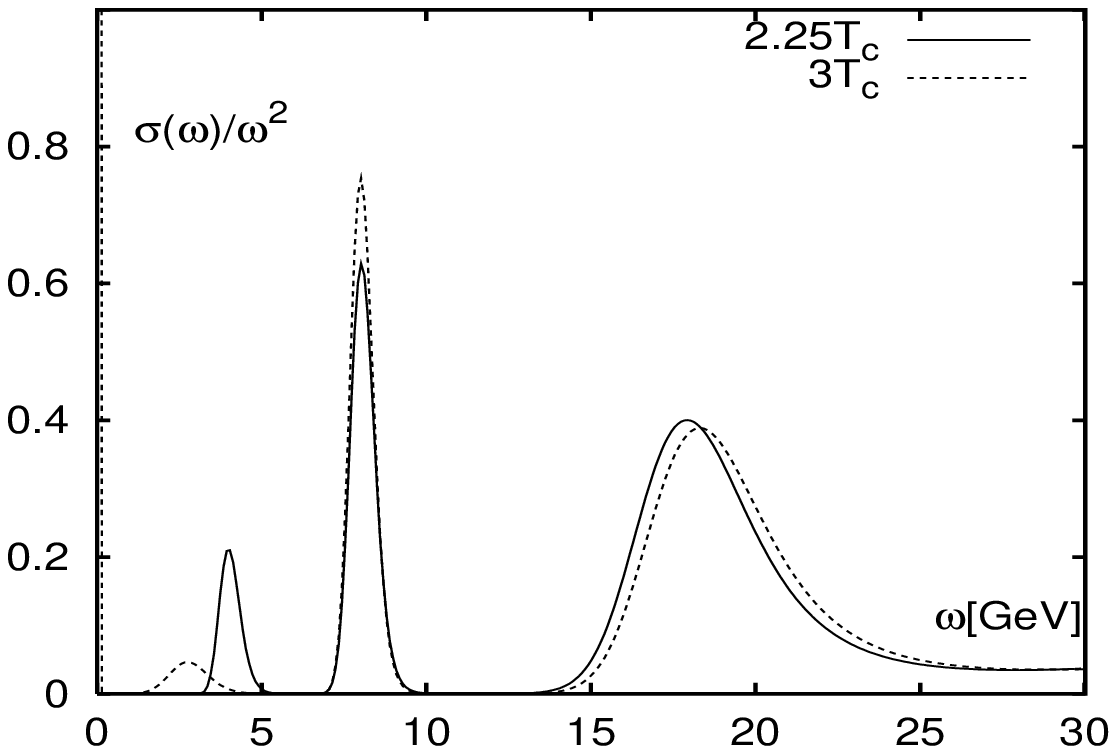,width=105mm}}
\put(-1.7,-1.0){\epsfig{bbllx=0,bblly=0,bburx=564,bbury=742,
file=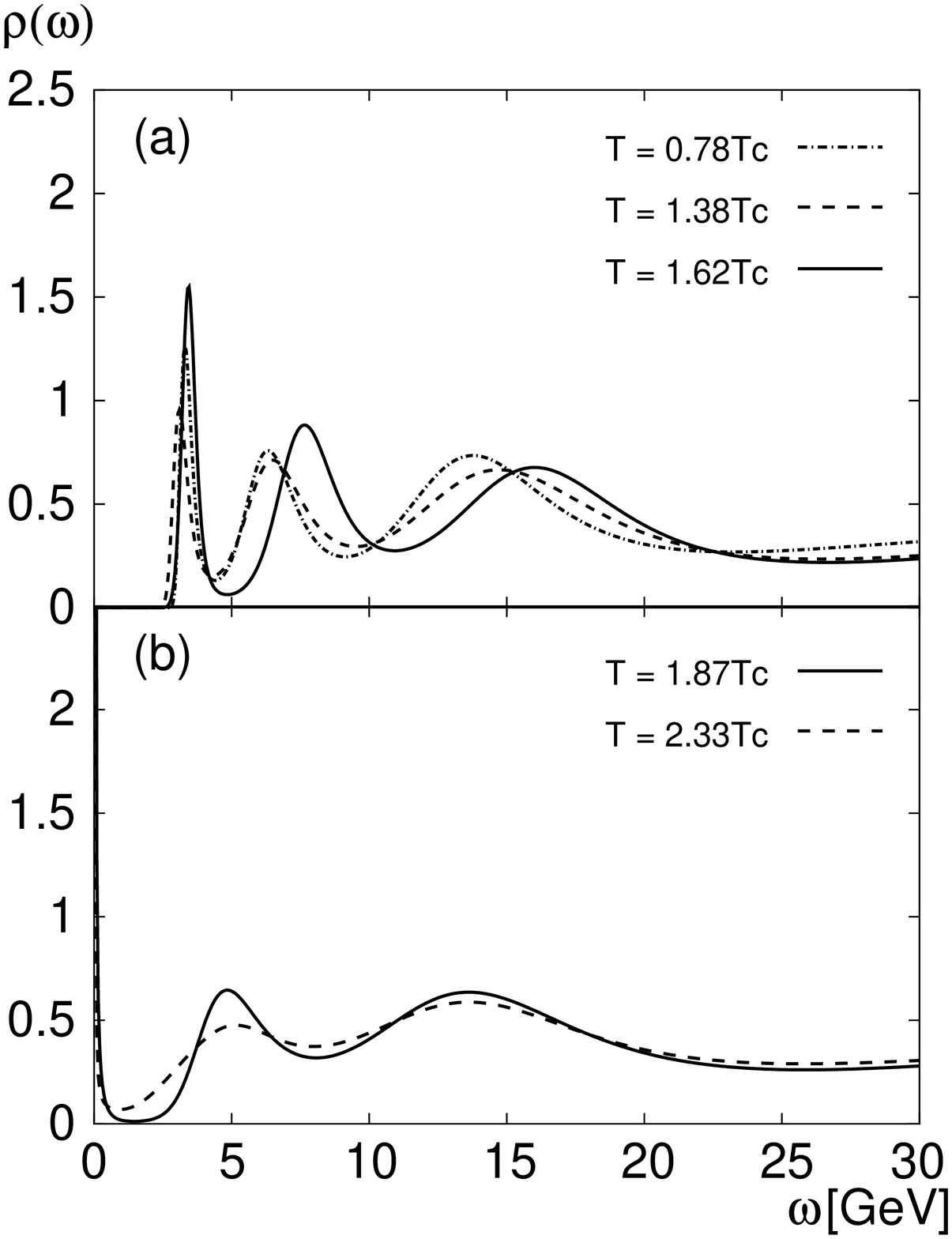,width=65mm}}
\put(-0.7,2.8){\epsfig{file=whitebox2.eps,width=0.6cm,height=0.4cm}}
\put(-0.7,6.3){\epsfig{file=whitebox2.eps,width=0.6cm,height=0.4cm}}
\put(-1.45,7.1){\epsfig{file=whitebox2.eps,width=0.7cm,height=0.4cm}}
\unboldmath
\end{picture}
\end{center}
\vspace*{0.5cm}
\caption{\label{spectral}Thermal vector spectral functions ($J/\psi$) 
from MEM analyses of meson correlation functions calculated in quenched
QCD on anisotropic \cite{Asakawa1} (left) and isotropic \cite{Datta1}
(right) lattices. The difference in scale on the ordinates of both figures
is due to different normalizations used for the hadronic currents.}
\end{figure}

The reconstruction of charmonium spectral functions has been performed 
using technically different approaches, e.g. using point-like
\cite{Asakawa1,Datta1} or smeared \cite{Umeda2} hadron sources on
isotropic \cite{Datta1} 
or anisotropic \cite{Asakawa1,Umeda1} lattice with standard Wilson
fermion \cite{Asakawa1} or improved Wilson fermion \cite{Datta1,Umeda1} actions.
The different calculations agree to the extent that no significant
modification of $S$-state spectral functions is observed up to
$1.5\; T_c$, {\it i.e.} the $J/\psi$ survives as a narrow bound state with
unchanged mass up to this temperature. This is shown in the upper part of
\Fref{spectral}. Current lattice calculations\footnote{Note that current
lattice calculations of spectral functions are strongly influenced by
lattice cut-off effects which show up most strongly at large energies,
$\omega$. In fact, only the first, low energy peak in the spectral functions 
shown in \Fref{spectral} is physical and insensitive to changes of the 
lattice cut-off. The other two peaks have been shown to be lattice
artifacts arising from ''Wilson doublers'' \cite{CPPACS}.} 
differ, however, on the 
structure of spectral functions for larger temperatures (lower part of 
\Fref{spectral}). While it is concluded in \cite{Asakawa1} that
the $J/\psi$ resonance disappears quite abruptly at $T\simeq 1.9\; T_c$
the analysis of \cite{Datta1} suggests that the resonance disappears
gradually; a resonance peak with reduced strength is still visible at 
$T= 2.25\; T_c$ and finally disappears completely at $T= 3\; T_c$.

To get control over the detailed pattern of dissolution of 
the heavy quark resonances clearly requires more refined studies. It also should
be noted that so far all existing lattice studies have been performed
in the quenched approximation. Although virtual quark loops are not 
expected to modify the qualitative picture obtained from these calculations
in the heavy quark sector of QCD one clearly has to understand their influence
on a quantitative level.  

\section{Conclusions}

Studies of spectral functions of charmonium states suggest that the $\bar{c}c$
ground states, $J/\psi$ and $\eta_c$, exist as well localized resonances
with masses essentially equal to their vacuum values at least up to temperatures
$T\;\simeq \; 1.5\; T_c$. Whether these states disappear abruptly at higher
temperatures or whether their contribution to the spectral functions
gradually weakens requires further studies.

We also discussed the reanalysis of heavy quark free energies and 
showed that these cannot directly be identified with the potential 
energy of a static quark anti-quark pair. 
The elimination of entropy contributions to the free energy leads to
(potential) energies which generally have a larger dissociation 
energy for a $\bar{c}c$-pair. Taking this into account in   
potential model calculations, which in the past suggested a rather early
dissociation of heavy quark bound states, can well lead to dissociation 
temperatures consistent with results obtained from the analysis 
of spectral functions. 

\section*{Acknowledgments}
I would like to thank all my collaborators at Bielefeld, BNL and Swansea
that contributed to the various research projects that led to 
many of the results discussed here. In particular, I would like to thank
Saumen Datta, Peter Petreczky and Felix Zantow for their critical 
comments on this write-up.
This work has been supported by the German ministry for education and 
research (BmBF) under contract no. 06BI106 and the German research
foundation (DFG) under grant KA 1198/6-4.

\end{document}